\documentclass[%
 reprint,
superscriptaddress,
 amsmath,amssymb,
 aps,
prl,
]{revtex4-1}
\usepackage{etoolbox}
\usepackage{graphicx}
\usepackage{dcolumn}
\usepackage{bm}
\usepackage{braket}
\usepackage{float}
\usepackage{xcolor}
\usepackage{comment}
\usepackage{hyperref}
\hypersetup{
    colorlinks=true,
    linkcolor=blue,
    citecolor=blue,
    filecolor=blue,
    urlcolor=blue,
    breaklinks=true  
}

\begin{document}

\preprint{APS/123-QED}

\title{Population-resolved measurement of an avoided crossing of light-dressed states}
\author{Noah~Schlossberger}
\email{noah.schlossberger@nist.gov}
\affiliation{National Institute of Standards and Technology, Boulder, Colorado 80305, USA}

\author{Nikunjkumar~Prajapati}
\affiliation{National Institute of Standards and Technology, Boulder, Colorado 80305, USA}

\author{Eric~B.~Norrgard}
\affiliation{National Institute of Standards and Technology, Gaithersburg, Maryland 20899, USA}
\author{Stephen~P.~Eckel}
\affiliation{National Institute of Standards and Technology, Gaithersburg, Maryland 20899, USA}

\author{Christopher~L.~Holloway}
\affiliation{National Institute of Standards and Technology, Boulder, Colorado 80305, USA}
\date{\today}

\begin{abstract}
A two-level system coupled by a coherent field is a ubiquitous system in atomic and molecular physics. In the rotating wave approximation, the light-dressed states are well described by a simple 2$\times$2 Hamiltonian which can be easily solved analytically and is thus used in quantum mechanics education and as a basis for intuition for more complicated systems. The solution to the Hamiltonian is an avoided crossing between the light-dressed ground and excited states. In experiments, the avoided crossing is probed spectroscopically, meaning only the energies, or eigenvalues of the Hamiltonian, are measured. Here, we present a measurement of the avoided crossing which also resolves population, thus indicating the amplitude coefficients of the \textit{eigenvectors} of the Hamiltonian. We perform the measurement in Rydberg states of cold rubidium atoms, resolving the energies spectroscopically with our pump lasers and the populations of each state using selective field ionization.
\end{abstract}

\maketitle

\section{Introduction}
A two-level system interacting with a near-resonant coherent field serves as a foundational toy model in quantum optics and atomic physics. Despite its simplicity, it captures essential physics such as Rabi oscillations, spectroscopic lineshapes, and the emergence of light-dressed states. When the system is analyzed in the rotating wave approximation, it reduces to a $2\times2$ Hamiltonian that can be diagonalized analytically, yielding a pair of eigenstates whose energies exhibit an avoided crossing as the detuning is varied. This avoided crossing, known as the Autler-Townes doublet \cite{PhysRev.100.703}, provides intuitive insight into more complex multi-level or strongly coupled systems. As such, the model plays a key role not only in pedagogy but also in practical applications, such as in  quantum computing \cite{Li2012} and memory \cite{Saglamyurek2018} as well as SI-traceable metrology of radio-frequency fields \cite{Schlossberger2024}.

The avoided crossing associated with the Autler-Townes effect has been observed and mapped in a number of different physical systems including atoms \cite{GRAY1978359}, molecules \cite{Schmitz2015}, nitrogen vacancy centers \cite{Zhou2017}, and superconducting Josephson junctions \cite{PhysRevLett.103.193601, Suri_2013}. However, the effect is typically measured spectroscopically. While these measurements probe the energy of the light-dressed states and thus the eigenvalues of the Hamiltonian, they do not resolve the eigenvectors. By performing spectroscopy of the avoided crossing with a state readout that resolves population in each of the two coupled states, we present a more complete experimental characterization of the avoided crossing which includes information about the eigenstates of the Hamiltonian. In this paper we map the energies of the avoided crossing as in previous measurements, but also determine the projection of the dressed states to the atomic states at each point.

\section{Hamiltonian and theory}
A two-level system interacting with a resonant light field can be solved using the rotating frame approximation. If the atom's ground state $\ket{g}$ and excited state $\ket{e}$ are coupled by radiation with a Rabi rate $\Omega$ at a detuning $\Delta$ from resonance, then in the semi-classical dressed-state basis the Hamiltonian $\hat H$ is given by \cite{PrinciplesLaserSpec}
\begin{equation}
\hat H = -\hbar \Delta \ket{e}\bra{e} + \frac{\hbar \Omega}{2}\left(\ket{e}\bra{g} + \ket{g}\bra{e}\right),
\end{equation}
where $\hbar$ is the reduced Planck's constant, which has eigenvalues (energies)
\begin{equation}
    E_\pm = - \frac{\hbar \Delta}{2} \pm \frac{\hbar \sqrt{\Omega^2 + \Delta^2}}{2} \label{eq:evals}
\end{equation}
and eigenvectors (eigenstates)
\begin{equation}
\ket{\pm} = \frac{\left(\frac{\Delta  \raisebox{0.2ex}{$\pm$} \sqrt{\Delta ^2+\Omega ^2}}{\Omega}\right)\ket{g} + \ket{e}}{\sqrt{\left| \frac{\Delta \raisebox{0.2ex}{$\pm$} \sqrt{\Delta ^2+\Omega ^2} }{\Omega }\right| ^2+|1|^2}}. \label{eq:evecs}
\end{equation}

\begin{figure}
    \includegraphics[scale = 0.8]{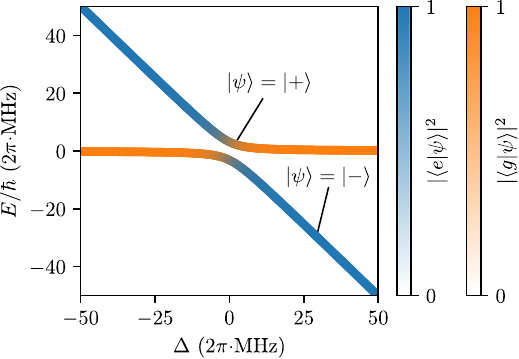}
    \caption{The avoided crossing described by equations \ref{eq:evals} and \ref{eq:evecs} for a Rabi rate of  $\Omega = 2\pi \cdot 1$ MHz. The color represents the projection of each state onto the atomic basis states.}
    \label{fig:basictheory}
\end{figure}

These eigenstates and eigen-energies are shown as a function of the detuning $\Delta$ in Fig. \ref{fig:basictheory}, with the y-axis representing the eigenvalues and the color representing the component of the eigenvectors in each of the basis states.

The features of note are that on resonance ($\Delta = 0$), both eigenstates are a 50:50 superposition of $\ket{e}$ and $\ket{g}$, and they are separated in energy by exactly $\hbar \Omega$.

\section{Measurement}
Here, we measure an avoided crossing between two highly excited states of $^{85}$Rb:
\begin{equation}
    \begin{cases}
    \ket{g} = \ket{32S_{1/2}} \\
    \ket{e} = \ket{32P_{3/2}}.
    \end{cases}
\end{equation}

\begin{figure}
\includegraphics[scale = 0.9]{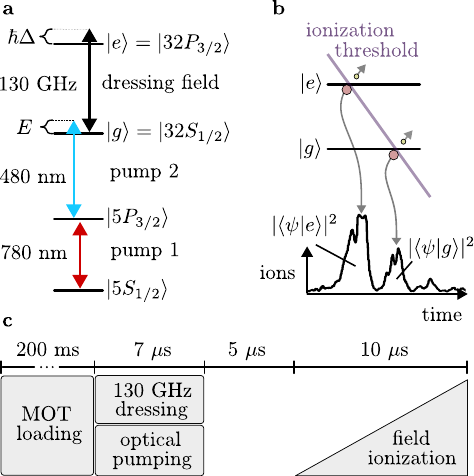}
\caption{Experimental scheme for measuring the dressed states. \textbf{a} The energy level scheme for the optical pumping and light dressing. \textbf{b} The state readout scheme. As the electric field is ramped, the states ionize at different times, resulting in an ionization signal with peaks for each state. \textbf{c} The timing of the experiment.}
\label{fig:experimentscheme}
\end{figure}

The experimental scheme is presented in Fig. \ref{fig:experimentscheme}. Each instance of the measurement begins with roughly $10^6$ $^{85}$Rb atoms with a temperature of roughly 1 mK in a magneto-optical trap (MOT) described in \cite{PhysRevResearch.7.L012020}. Two lasers, pump 1 and pump 2, are then applied to populate $\ket{g}$ (Fig. \ref{fig:experimentscheme}a).  Simultaneously, the $\ket{g}$ and $\ket{e}$ states are dressed by applying a roughly 130 GHz radio-frequency (RF) field with detuning $\Delta$.  
The energy of the dressed states is measured by scanning the the energy of the second photon (pump 2), as population is only transferred into the dressed state when the second photon is on resonance. 
\begin{figure*}
    \includegraphics[scale = 0.9, trim = {0 0 0.5cm 0}, clip]{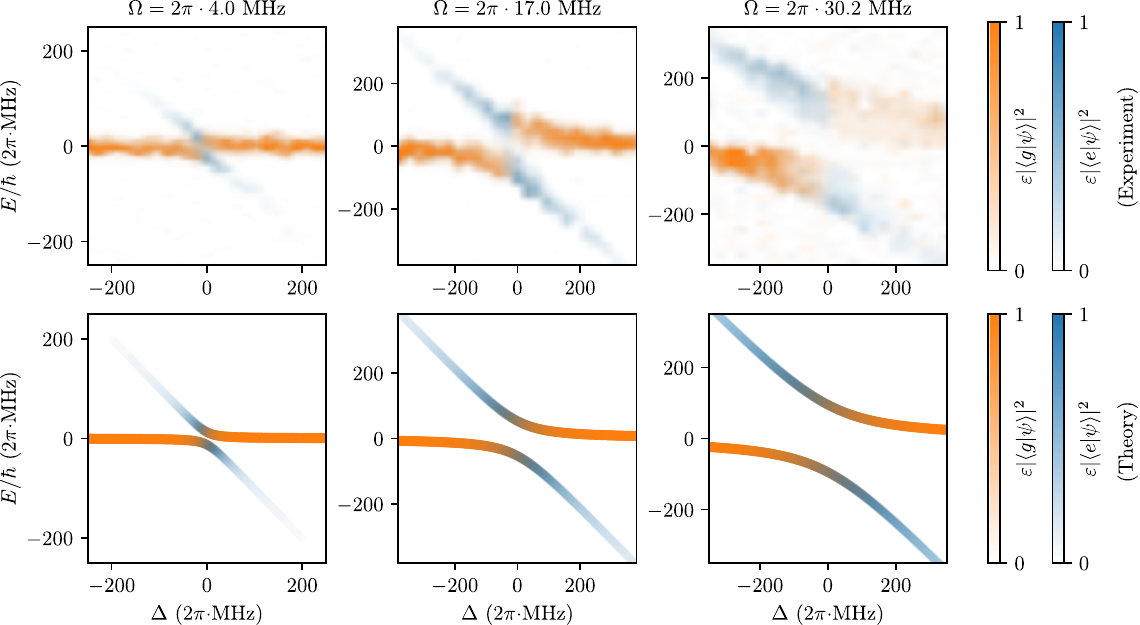}
    \caption{Experimental (top) and theoretical (bottom) mappings of the avoided crossings at various RF powers. The theoretical curves are calculated from Eqs. \ref{eq:evals} and \ref{eq:evecs} using a Rabi frequency inferred from a measured calibration between the square root of the power emitted by the source and the Autler-Townes splitting observed.}
    \label{fig:all_expdata}
\end{figure*}\\

We then read out state populations by performing selective field ionization \cite{gallagher1994rydberg, PhysRevA.51.4010, PhysRevResearch.7.L012020} (Fig. \ref{fig:experimentscheme}b). Because the states $\ket{g}$ and $\ket{e}$ are highly excited, they may be ionized simply by applying an electric field on the order of 60 kV/m. Because the state $\ket{e}$ ionizes at a lower field than $\ket{g}$, ramping the electric field ionizes the two states at different times. We collect the ions with a channel electron multiplier and obtain two peaks corresponding to the two quantum states $\ket{g}$ and $\ket{e}$.  The time-integral of these peaks is proportional to the number of atoms in each state. More details on the selective field ionization setup can be found in \cite{PhysRevResearch.7.L012020}.  The timing of the experimental sequence is shown in Fig. \ref{fig:experimentscheme}c.

Because we are populating the atoms via a laser excitation to the ground state, the population transfer efficiency $\varepsilon$ of the pumping lasers to populate a state $\ket{\psi}$ will go as 
\begin{equation}
   \varepsilon =  |\langle\psi|g\rangle|^2,
\end{equation}
and the integrals of the two ionization peaks represent measurements of $\varepsilon|\langle\psi|g\rangle|^2$ and $\varepsilon|\langle\psi|e\rangle|^2$.

\section{Results}

To map the avoided crossing of Eq.\,\ref{eq:evals} as in Fig. 1, we slowly scan the frequency of the pump 2 laser as we repeat the sequence described in Fig. \ref{fig:experimentscheme}c. For each shot, we integrate the two peaks in the ionization signal to get the state populations. This scan represents a vertical line of the avoided crossing map. We then change the frequency of the 130~GHz source and repeat this sequence to map out the spectrum for various $\Delta$. The results are shown in Fig. \ref{fig:all_expdata}.   We find that our measurements of both the eignenvalues and eignvector amplitudes are in excellent agreement with the predictions of Eq.\,\ref{eq:evals} and Eq.\,\ref{eq:evecs}, respectively, for a wide range of $\Delta$.

The Rabi rates used in the theory are calculated from the power supplied by the 130~GHz source, with a calibration provided by a linear fit of the Autler-Townes splitting versus the square root of the power applied. The broadening of the states observed for large Rabi rates is most likely due to inhomogeneity of the 130~GHz field, as well as the contribution of different hyperfine $F$ and associated projection $m_F$ states, which have different transition dipole moments and therefore have different Rabi rates for the same field strength. Furthermore, although all photons are nominally $\pi$-polarized, reflections of the RF inside the vacuum chamber can scramble the polarization, thus further increasing the number of contributing dipole moments.

\section{Landau-Zener}
\begin{figure}
\includegraphics[scale = 0.9]{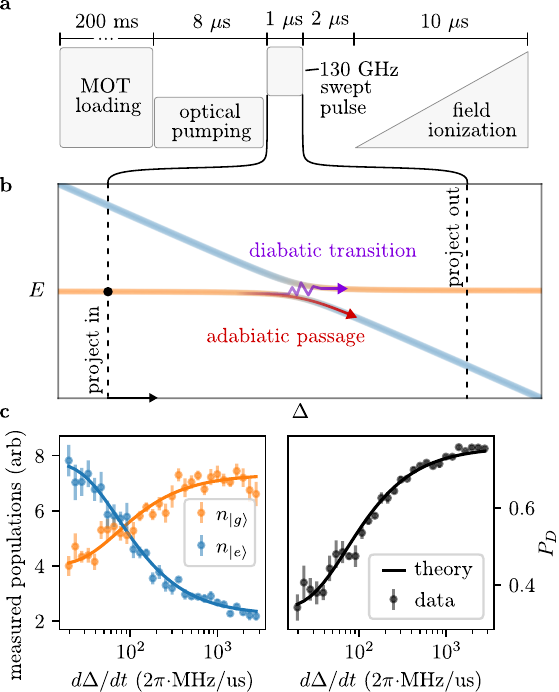}
\caption{Landau-Zener transition rate mapping. \textbf{a} The timing of the experiment. \textbf{b} The transition in the dressed state picture. \textbf{c} Experimental mapping of the populations and the associated diabatic transition probability $P_D$ for various sweep rates. Each point represents 10 measurements. Here the solid lines represent Landau-Zener theory with a fit offset and contrast (Eq. \ref{eq:LZfit}).}
\label{fig:LandauZener}
\end{figure}

To further make use of the readout capability of selective field ionization, we map out the Landau-Zener transition probability as in \cite{PhysRevLett.104.133003, PhysRevA.51.646, PhysRevLett.118.257701, PhysRevB.90.100201}, but with the ability to simultaneously observe $\ket{g}$ and $\ket{e}$ state populations. To do this, we first prepare the atoms in state $\ket{g}$ with the pump lasers in the absence of the RF dressing field, then pulse the RF field and sweep it through resonance, and then finally readout the populations using selective field ionization (Fig. \ref{fig:LandauZener}a). When we first turn on the RF at a large negative detuning, the atoms are projected from the $\ket{g}$ state into the $\ket{-}$ state (Fig. \ref{fig:LandauZener}b). If the RF is slowly swept through resonance, the state adiabatically tracks and remains in the $\ket{-}$ state, which then projects into the $\ket{e}$ state at large positive detuning when we turn the RF off. If the RF is rapidly swept through resonance, the state  diabatically transitions to the $\ket{+}$ state, and thus  projects into the $\ket{g}$ state at large positive detuning. The probability of this diabatic transition $P_D$ is given by the Landau-Zener equation \cite{1370290617742844819}: 
\begin{equation}
    P_{D} = \exp\left(-2\pi\frac{ (\Omega/2)^2}{\left|\frac{d}{dt}( \Delta)\right|}\right).
    \label{eq:Pd}
\end{equation}
Here, $\frac{d}{dt}( \Delta)$ represents the sweep rate of the RF detuning.

Our experimental results are shown in Fig. \ref{fig:LandauZener}c. We measure the populations after the RF sweep using selective field ionization at various RF sweep rates $d\Delta/dt$. From these populations we infer the diabatic transition fraction as

\begin{equation}
 P_{D} =    \frac{|\langle\psi|g\rangle|^2}{ |\langle\psi|g\rangle|^2  +   |\langle\psi|e\rangle|^2}.
\end{equation}
We then compare the populations and diabatic transition fraction to Eq. \ref{eq:Pd}. We determine the Rabi rate $\Omega$ spectroscopically as in the previous measurements using the Auter-Townes splitting, finding the value to be $2\pi\cdot 0.54$~MHz (set to fully map out $P_D$ within the sweep capabilities of the RF source). Because blackbody radiation couples the two states during the measurement, the population contrast is reduced. We then float an offset $A$ and a contrast $B$ when comparing theory to data, and fit our data (solid lines in Fig. \ref{fig:LandauZener}c) to
\begin{equation}
    A + B\cdot  P_D(d\Delta/dt) \label{eq:LZfit}.
\end{equation}
With the shape and timescale of the curve $P_D$ fixed and only the offset and contrast fit, the curves match the experimentally measured data within the statistical uncertainty.

\section{Conclusion}
With the ability to resolve populations in both states simultaneously using selective field ionization, we have mapped out an avoided crossing of two Rydberg states in the presence of a radio frequency field coupling the two states together. This allows us to see the amplitude coefficients of the eigenstates of the dressed state Hamiltonian, providing a closer look at the classic picture of a coupled two-level system. We have also measured populations as we scanned across the resonance, demonstrating the crossover between diabatic state transition and adiabatic passage, with the ability to see population leaving the ground state and entering the excited state simultaneously.
Selective field ionization can therefore serve as a useful tool for characterizing the $2\times 2$ eigensystem of Rydberg atoms in the presence of an RF dressing field. 

\section{Methods}
Pump 1 was locked to the D2 transition in a vapor cell of rubidium using saturated absorption spectroscopy. The energy detuning $E$ of pump 2 was referenced to another Rubidium vapor cell using two-photon electromagnetically induced transparency, and the magnitude of the frequency scan was calibrated using a wavemeter.

Pump 1 had a power of 330~$\mu$W and a beam size at a full-width at half maximum (FWHM) of 530~$\mu$m. Pump 2 had a power of 10~mW and a beam size at FWHM of 650~$\mu$m. The 130~GHz dressing field was applied from a WR8 horn with a directivity of 23~dBi through a vacuum viewport with a 35~mm aperture a distance of 25~cm away from the center of the MOT. Both pump lasers and the 130~GHz field were all linearly polarized in the same axis. However, reflections of the microwave radiation inside the vacuum chamber likely scrambled the polarization at the location of the atoms.

During the Landau-Zener measurements, the frequency of the 130~GHz field was swept with a sinusoidal frequency modulation centered around zero detuning, with a modulation rate of 0.5~MHz such that the frequency detuning was swept from negative to positive in the 1~$\mu$s pulse. We took $d\Delta/dt$ to be the frequency derivative at the zero crossing. The frequency modulation rate was kept constant, and $d\Delta/dt$ was varied by changing the depth of the frequency modulation.

Error bars represent statistical uncertainty, calculated as the standard deviation divided by the square root of the number of measurements.

\subsection*{Acknowledgments}
This work was completed on equipment funded by the National Institute of Standards and Technology (NIST) through the Innovations in Measurement Science (IMS) program, and was partially funded by the NIST-on-a-Chip (NOAC) program. The authors thank Dazhen Gu for loaning a frequency extension head to produce the 130~GHz radiation.
\subsection*{Conflict of Interest}
\vspace{-3mm}
The authors have no conflicts to disclose.

\vspace{-3mm}
\subsection*{Data Availability Statement}
\vspace{-3mm}
All data presented in this paper is available at \href{https://doi.org/10.18434/mds2-3845}{https://doi.org/10.18434/mds2-3845}.


%
\end{document}